\documentclass[conference]{IEEEtran}
\IEEEoverridecommandlockouts
\usepackage{cite}
\usepackage{amsmath,amssymb,amsfonts}
\usepackage{algorithmic}
\usepackage{graphicx}
\usepackage{tabularx}
\usepackage{textcomp}
\usepackage{bm}
\usepackage{float}
\usepackage{multirow}
\usepackage{booktabs}
\usepackage{xcolor}
\def\BibTeX{{\rm B\kern-.05em{\sc i\kern-.025em b}\kern-.08em
    T\kern-.1667em\lower.7ex\hbox{E}\kern-.125emX}}

\makeatletter
\newcommand{\linebreakand}{%
  \end{@IEEEauthorhalign}
  \hfill\mbox{}\par
  \mbox{}\hfill\begin{@IEEEauthorhalign}
}
\makeatother

\begin{document}

\title{$T^3$: Multi-level Tree-based Automatic Program Repair with Large Language Models

}
\author{
\IEEEauthorblockN{1\textsuperscript{st} Quanming Liu}
\IEEEauthorblockA{\textit{School of Computer and Information Technology} \\
\textit{Shanxi University} \\
Taiyuan, China \\
liuqm@sxu.edu.cn}
\and
\IEEEauthorblockN{2\textsuperscript{nd} Xupeng Bu}
\IEEEauthorblockA{\textit{School of Computer and Information Technology} \\
\textit{Shanxi University} \\
Taiyuan, China \\
202322407003@email.sxu.edu.cn}
\linebreakand
\IEEEauthorblockN{3\textsuperscript{rd} Zhichao Yan}
\IEEEauthorblockA{\textit{School of Computer and Information Technology} \\
\textit{Shanxi University} \\
Taiyuan, China \\
202312407023@email.sxu.edu.cn}
\and
\IEEEauthorblockN{4\textsuperscript{th} Ru Li}
\IEEEauthorblockA{\textit{School of Computer and Information Technology} \\
\textit{Shanxi University} \\
Taiyuan, China \\
liru@sxu.edu.cn}
}
\maketitle

\begin{abstract}
Automatic Program Repair (APR) is a core technology in software development and maintenance, with aims to enable automated defect repair with minimal human intervention. In recent years, the substantial advancements in Large Language Models (LLMs) and the Chain-of-Thought (CoT) techniques have significantly enhanced the reasoning capabilities of these models.
However, due to the complex logic and multi-step reasoning ability needed, the application of CoT techniques in the APR domain remains insufficient. This study systematically evaluates the performance of several common CoT techniques in APR tasks and proposes an innovative framework $T^3$, which integrates the powerful reasoning capabilities of LLMs with tree search, effectively improving the precision of generating candidate repair solutions. Furthermore, $T^3$ provides valuable guidance for optimizing sample selection and repair strategies in APR tasks, establishing a robust framework for achieving efficient automated debugging.
\end{abstract}


\section{Introduction}
Large Language Models (LLMs) have demonstrated exceptional performance due to their large number of model parameters and rich training data \cite{zhang2024systematic, yan2024atomic}. 
They have achieved significant results in numerous fields related to source code, including code generation, summarization, and test generation, making a profound influence on research methods and paradigms in the field of software engineering.

\begin{figure}[t]
\centerline{\includegraphics[width=0.5\textwidth]{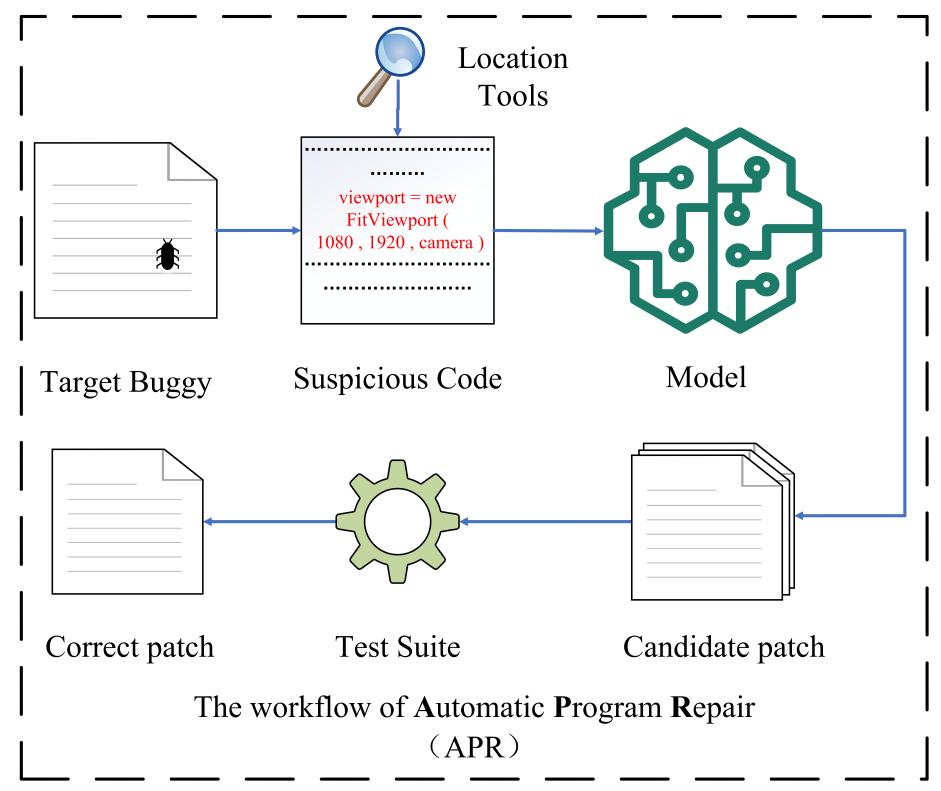}}
\caption{The general process of Automated Program Repair includes the following steps: first, locating the fault in the problematic program, then generating candidate repair patches. Finally, the correct repair patch is confirmed through test case validation.}
\label{fig:introduction}
\end{figure}

With the deep application of LLMs in the field of software engineering, their impact is increasingly significant. However, as software continues to grow in size and complexity, the number of software vulnerabilities is also increasing rapidly. These vulnerabilities may not only cause inconvenience to users but also cause huge economic losses. Especially in key areas such as medical care and transportation, the existence of vulnerabilities may lead to serious security risks and immeasurable economic damage. As a result, developers are compelled to allocate substantial time and resources to manually identify and fix these issues.
Automated Program Repair (APR) technology has emerged, as illustrated in Fig.~\ref{fig:introduction}, aiming to transform labor-intensive manual maintenance tasks into precise and efficient automated repair processes\cite{yiheng2024nmt}.

Existing LLMs-based APR technologies can be broadly divided into three types: Fine-tuning-based repair methods, Cloze-style repair methods, and Conversation-based repair methods.
Fine-tuning-based repair methods: RAP-GEN \cite{wang2023rap} and Mashhadi \cite{mashhadi2021applying} enhance repair performance by fine-tuning pre-trained models like CodeT5 \cite{wang-etal-2023-codet5,wang-etal-2021-codet5} and CodeBERT \cite{feng-etal-2020-codebert} on target datasets. Although effective within the scope of the training data, these methods demonstrate constrained generalization capabilities, with repair performance significantly declining for programs outside the target datasets.
Cloze-style repair methods: ALPHA \cite{xia2022less} and GAMMA \cite{zhang2023gamma} focus on masking erroneous code segments and predicting the correct content for masked positions. While these methods are dependent on precise error localization, they lack transparency in the reasoning process behind patch generation, making them less interpretable.
Conversation-based repair methods, such as ChatRepair \cite{Xia2024}, employs an iterative process to generate repair patches and validate feedback through conversational exchanges. Although the circular improvement approach enhances repair capabilities, it is difficult to address those problems with highly complex and ambiguous. These limitations hinder the generation of innovative solutions, increase the risk of entering infinite feedback loops, and ultimately constrain further advancements in repair efficiency.

To address these limitations, we propose $T^3$, a Multi-level Tree-based Automatic Program Repair framework utilizing Large Language Models. $T^3$ decomposes the reasoning process for generating candidate patches into four stages: sample retrieval, cause analysis, repair plan generation, and patch generation. In the cause analysis and repair plan generation stages, the Forest of Thinking approach enhances reasoning by constructing multiple parallel reasoning trees. The top-n results, selected based on cross-tree consistency and voting mechanisms, are then used for further processing.

Furthermore, $T^3$ overcomes the constraints of single-path reasoning inherent in traditional chain-based methods such as CoT \cite{wang2023rap}, Tree-of-Thought \cite{tree-of-thought-prompting}, Plan-and-Solve \cite{wang-etal-2023-plan}, and Analogical Reasoning \cite{yasunaga2024large}. These methods rely on a single reasoning chain, limiting their ability to handle complex problems. By employing a multi-level tree-based reasoning structure, $T^3$ enhances diversity in reasoning paths, significantly improving both reasoning capabilities and problem-solving effectiveness.

Our contributions are summarized as follows:
\begin{itemize}
\item This paper proposes a new reasoning framework $T^3$ (Multi-level Tree-based Automatic Program Repair with Large Language Models), which can enhance path diversity and overall repair capability in the patch generation process.
\item This paper proposes a novel result selection method that improves the diversity of selected results while ensuring the stability of the results. The method achieves this by balancing the variability of the results with consistency, thus preserving the overall reliability of the selection process.
\item This paper introduces a novel repair method that achieves substantial improvements over previous approaches; specifically, it improves repair accuracy by 11.20\% and 9.90\% compared to methods based on CoT reasoning.
\end{itemize}

\section{RELATED WORK}

\subsection{Auto Program Repair}

In software development, addressing software defects effectively is critical to enhancing developer productivity and reducing the manual effort required to fix an ever-growing number of issues. APR technology has long aimed to eliminate software defects and has evolved significantly over the years. It encompasses various approaches, including search-based methods\cite{sun2018search}, constraint-based methods\cite{yi2022speeding}, template-based methods\cite{liu2019tbar}, and learning-based methods\cite{li2022dear}. Among these, template-based and learning-based techniques have demonstrated superior performance.
Template-based methods rely on predefined templates created by experts. These templates match specific patterns in the target program and generate corresponding repair strategies. On the other hand, learning-based methods leverage machine learning and deep learning techniques to reframe program repair as a translation task.
The Neural Machine Translation (NMT) paradigm is widely adopted in APR, treating the process of fixing defective code as analogous to translating it into corrected code. By applying Natural Language Processing (NLP) NMT models, this approach learns the transformation from defective to repaired statements\cite{yiheng2024nmt}.
With the advent of large pre-trained models, researchers have explored using these models directly for patch prediction. One such approach involves treating program repair as a cloze-style task, where the defective code sections are masked, and the model predicts the correct replacements\cite{xia2022less,zhang2023gamma}.
Another emerging approach is conversation-based repair\cite{xia2023conversational}, which utilizes complex prompts containing diverse, valuable information, such as defective code snippets, troubleshooting processes, and even execution feedback\cite{Xia2024}. By interacting with LLMs, this method generates accurate patches dynamically and contextually.
\subsection{Large language Models}
Large language models (LLMs) are a disruptive and innovative technology in the field of Natural Language Processing. They widely adopt the Transformer architecture and are trained on vast text datasets, demonstrating exceptional processing capabilities. Due to their training on large-scale data, LLMs have not only demonstrated exceptional reasoning capabilities in complex and dynamic reasoning tasks \cite{wang2024large,wang2024made,wang2024ime}, but have also achieved notable advancements in various code-related applications, showcasing their remarkable generalization abilities \cite{chen2021evaluating,fried2022incoder,xu2022systematic}. 
When applied to downstream tasks, researchers typically fine-tune pre-trained models to enhance their performance on specific tasks. This approach optimizes the model based on the task-specific data. However, in recent years, prompting techniques have emerged as a new method to enhance the model's reasoning ability for downstream tasks without the need for fine-tuning. The core idea of prompting is to design an appropriate task description and provide a few examples to help the model understand the task requirements and objectives \cite{reynolds2021prompt}. In this way, the model can quickly adapt and effectively perform new tasks without additional training or fine-tuning.
Prompting not only simplifies the application process of models but also provides new insights into the efficient reasoning of LLMs in practical applications. As a result, models can exhibit good performance on many real-world tasks with only a small number of prompts\cite{sahoo2024systematic}.
\subsection{Chain of Thought}
The emergence of CoT reasoning has introduced a concise and effective approach that unlocks the reasoning potential of LLMs. By embedding structured chains of thought directly into prompts, this method enables LLMs to perform complex reasoning tasks. Researchers have shown that using a triplet structure—comprising input, thought chain, and output—in prompts allows models to demonstrate strong reasoning capabilities with minimal examples.
Since Wei's introduction of the CoT concept in 2022 \cite{wei2022chain}, this methodology has rapidly evolved, spawning a variety of influential variants: ``Let’s Think Step by Step" \cite{kojima2022large} advocates for systematic, step-by-step reasoning.
``Self-Consistency" \cite{wang2023selfconsistency} focuses on generating and refining self-consistent reasoning paths.
``Analogical Reasoning" \cite{yasunaga2024large} encourages the generation of examples analogous to the original problem, solving these examples first to inform the final solution.
``Plan-and-Solve" \cite{wang-etal-2023-plan} emphasizes problem decomposition followed by strategic resolution. ``Tree-of-Thoughts" \cite{tree-of-thought-prompting} introduces hierarchical reasoning to tackle complex tasks.
These prompting techniques highlight the broad applicability of CoT methods across various domains and their significant performance improvements. However, despite the substantial value CoT technologies have demonstrated in many areas, there remains a notable gap in research exploring their application in APR.
To address this, the study aims to thoroughly investigate how CoT techniques can enhance the reasoning capabilities of LLMs to improve their performance in APR tasks. Custom prompt templates based on four common CoT variants were designed, and experiments were conducted to comparatively analyze their effectiveness in this domain.
\section{PROPOSED METHOD}
\begin{figure}[htbp]
\centerline{\includegraphics[width=0.5\textwidth]{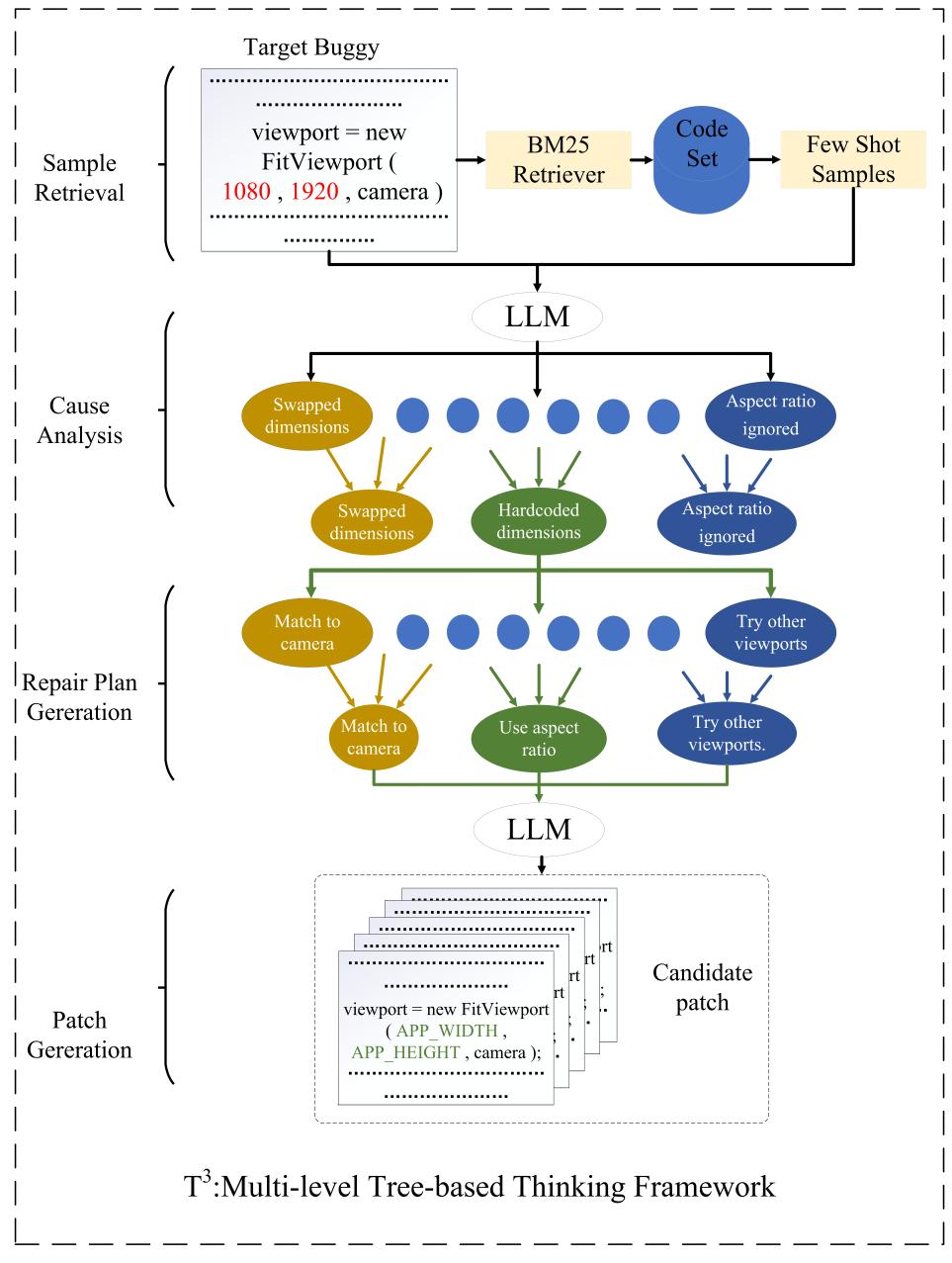}}
\caption{The $T^3$ method involves the following structured steps:
1.Sample Retrieval: Perform a keyword search utilizing the BM25 algorithm to identify relevant code examples from the codebase that are similar to the target buggy snippet.
2.Cause Analysis: Analyze the identified examples alongside the target buggy snippet to determine the underlying causes of the error.
3.Repair Plan Generation: Formulate a detailed repair strategy based on the previously analyzed error causes.
4.Patch Generation: Generate a corresponding program patch by implementing the repair plan to resolve the identified issues effectively.}
\label{$T^3$}
\end{figure}

This study proposes a novel reasoning framework, $T^3$, specifically designed to enhance the APR capabilities of LLMs. As illustrated in Fig.~\ref{$T^3$}, the framework comprises four key components: Sample Retrieval, Cause Analysis, Repair Plan Generation and Patch Generation.

\subsection{Sample Retrieval}
To retrieve code examples similar to the target program snippet \( P_{\text{target}} \), we employ the BM25 algorithm. This retrieval process is formalized as:  
\begin{equation}
C_{\text{retrieved}} = \text{BM25}(P_{\text{target}}, \mathcal{S}, k),
\end{equation}  
where \( \mathcal{S} \) denotes the code repository, and \( k \) specifies the number of code examples to retrieve. As described in \cite{ahmed2023better}, BM25 effectively identifies relevant examples by measuring their textual similarity to \( P_{\text{target}} \).  

After obtaining the relevant code examples \( C_{\text{retrieved}} \), the next step is to construct the input for the language model. This is done by concatenating the retrieved examples, the target program \( P_{\text{target}} \), and the identified error location \( L_{\text{target}} \):  
\begin{equation}
I_{\text{LLM}} = \text{concat}(C_{\text{retrieved}}, P_{\text{target}}, L_{\text{target}}).
\end{equation}  
The resulting composite input \( I_{\text{LLM}} \) integrates contextual information from similar examples, the target program, and the specific error location, providing the language model with comprehensive context for effective analysis.  

Finally, \( I_{\text{LLM}} \) is combined with a carefully designed prompt template and fed into a LLM. The LLM leverages the retrieved code examples and additional context to diagnose potential causes of the error. This structured approach enhances the LLM’s diagnostic capabilities by grounding its analysis in relevant code patterns and specific error-related information.  
\subsection{Cause Analysis}
At the cause analysis stage, we adopt the Forest of Thinking approach. Unlike the traditional Tree-of-Thought method, our approach constructs a forest consisting of multiple independent yet interconnected reasoning trees. Each tree represents a distinct pathway for analyzing potential error causes. This enables the model to explore multiple hypotheses in parallel, thereby enhancing the comprehensiveness and robustness of error diagnosis.

All reasoning trees originate from the same initial problem statement but evolve independently as they traverse distinct reasoning trajectories. Specifically, we construct a set of $M$ independent error reasoning trees, where each tree adheres to its distinct logical deduction pathway:

\begin{equation} 
\mathcal{F} = \{T^1, T^2, \dots, T^M\}, 
\end{equation}
where each tree $T^i$ expands using the CoT approach to generate multiple candidate error causes:

\begin{equation} 
T^i = \{\text{CoT}(I_{\text{LLM}})\}_{j=1}^{N_i}. 
\end{equation}

Here, $N_i$ represents the number of candidate error causes generated within the $i$-th reasoning tree. Although all trees share a common starting point, they diverge in reasoning due to variations in logical patterns, leading to a broader exploration of potential error causes.

To improve the reliability of error cause selection, we introduce a self-consistency mechanism. Unlike prior work \cite{ahmed2023better}, which applies self-consistency for patch selection, we leverage this mechanism to filter and rank error causes. By aggregating the results from different reasoning trees, we identify the most frequently occurring error causes across the forest. This reduces bias and enhances the robustness of our analysis.

For each candidate error cause $r \in \mathcal{F}$, we compute its frequency across all reasoning trees:

\begin{equation}
f(r) = \sum_{i=1}^{M} \sum_{j=1}^{N_i} \mathbb{I}(r_{ij} = r),
\end{equation}
where the indicator function $\mathbb{I}(r_{ij} = r)$ is defined as:

\begin{equation}
\mathbb{I}(r_{ij} = r) = 
\begin{cases} 
1, & \text{if } r_{ij} = r, \\ 
0, & \text{otherwise.} 
\end{cases}
\end{equation}

Based on the computed frequencies, we select the most frequently occurring error causes to form the Top-$n$ set:

\begin{equation}
R_{\text{top}} = \{r^{(1)}, r^{(2)}, \dots, r^{(n)}\},
\end{equation}
where the selected error causes are ranked in descending order of occurrence:

\begin{equation}
f(r^{(1)}) \geq f(r^{(2)}) \geq \dots \geq f(r^{(n)}) \geq f(r), \quad \forall r \notin R_{\text{top}}.
\end{equation}

By leveraging a reasoning forest instead of a single tree, our approach ensures a more comprehensive exploration of potential error causes. This structured methodology not only reduces the risk of overlooking critical issues but also provides a solid foundation for the subsequent repair planning stage, ultimately improving the accuracy and effectiveness of program repair.

\subsection{Repair Plan Generation}
Building upon the results of the cause analysis stage, we formulate targeted repair plans using the Forest of Thinking approach. Instead of relying on a single reasoning trajectory, we construct a reasoning forest in which each tree originates from a different error cause identified in the cause analysis stage. This parallel exploration of multiple repair pathways enhances both the robustness and diversity of the solutions generated.

Given the target program \( T \), the retrieved examples \( E \) from the Sample Retrieval stage, and the top-ranked error causes \( R_{\text{top}} \), we construct \( M \) independent repair reasoning trees:

\begin{equation}
\mathcal{P} = \{T^1, T^2, \dots, T^M\},
\end{equation}

where each tree \( T^i \) originates from a specific error cause \( r^i \in R_{\text{top}} \) and expands using the CoT reasoning mechanism to generate multiple candidate repair solutions:

\begin{equation}
T^i = \{\text{CoT}(T, E, r^i)\}_{j=1}^{N_i}.
\end{equation}

Here, \( N_i \) represents the number of candidate repair plans generated within the \( i \)-th reasoning tree. Since each tree starts from a distinct error cause, this structured approach ensures that diverse repair strategies are explored.

To identify the most promising repair solutions, we aggregate repair plans across all trees and compute the frequency of each unique candidate \( p \in \mathcal{P} \):

\begin{equation}
f(p) = \sum_{i=1}^{M} \sum_{j=1}^{N_i} \mathbb{I}(p_{ij} = p),
\end{equation}
where the indicator function \( \mathbb{I}(p_{ij} = p) \) is defined as:

\begin{equation}
\mathbb{I}(p_{ij} = p) = 
\begin{cases} 
1, & \text{if } p_{ij} = p, \\ 
0, & \text{otherwise.}
\end{cases}
\end{equation}

After computing the frequency of each repair plan, we select the top \( n \) solutions with the highest occurrence rates to form the final set:

\begin{equation}
P_{\text{top}} = \{p^{(1)}, p^{(2)}, \dots, p^{(n)}\}.
\end{equation}

By employing a reasoning forest where each tree is rooted in a distinct error cause, our method systematically explores multiple repair strategies while minimizing bias toward any single reasoning path. This structured approach improves the accuracy and diversity of repair plan selection, ultimately leading to more effective patch generation for the target program.
\subsection{Patch Generation}
To generate specific patches, we input the original program \( T \), the retrieved examples \( E \), and the selected repair plans \( P_{\text{top}} \) into the language model. Leveraging the CoT method, the language model synthesizes the information to produce a final patch, formalized as:  
\begin{equation}
    \text{Patch} = \text{CoT}(T, E, P_{\text{top}}).
\end{equation}  

This iterative process, which involves progressively identifying error causes, formulating repair plans, and generating the final patch, represents a significant advancement over conventional approaches. By systematically breaking down the problem into manageable steps and incorporating diverse reasoning paths, our method overcomes the limitations of traditional reasoning techniques.  

As a result, the generated patches are not only more precise but also more comprehensive, addressing the underlying issues in a structured and effective manner. This approach ensures a higher likelihood of producing accurate fixes, ultimately enhancing the reliability and quality of the repair process.

\section{Experimental setup}

\subsection{Dataset}
The MODIT dataset\cite{ahmed2023better} used in this study includes two subsets: B2Fs (with smaller code sequences) and B2Fm\cite{tufano2019empirical}. The MODIT dataset encompasses bug fix commits collected from GitHub, along with detailed commit logs. 
\subsection{Settings}
This study employs gpt-3.5-turbo and gpt-4o-mini \footnote{To ensure reproducibility, we use the version gpt-3.5-turbo-0125 and gpt-4o-mini-2024-07-18}, both accessed via OpenAI's API. Given that previous experiments conducted by Toufique Ahmed \cite{ahmed2023better} were based on the deprecated code-davinci-002 model, we uniformly select the widely used gpt-3.5-turbo and gpt-4o-mini to eliminate the impact of foundational model discrepancies on repair performance.To ensure the reliability of our results, each program in every experimental group is tested with 30 generated samples, and the temperature parameter is set to 0.7 to mitigate the influence of randomness.
\subsection{Metrics}
To evaluate the model's repair capabilities, this study employs two key metrics: repair rate and SC accuracy \cite{wang2023selfconsistency}.
\begin{equation}
R = \frac{N_{\text{correct}}}{N_{\text{total}}}
\end{equation}
Repair Rate:
the repair rate is defined as the proportion of successfully repaired programs to the total number of programs, reflecting the model's overall effectiveness in resolving defects. Specifically, $N_{\text{correct}}$ denotes the number of programs successfully repaired, while $N_{\text{total}}$ represents the total number of programs.
\par
\begin{equation}
\text{SC Accuracy} = \frac{N_{\text{selected-correct}}}{N_{\text{total-correct}}}
\end{equation}
SC accuracy:The sc accuracy evaluates the probability that the most frequently generated patch during the reasoning process is the correct patch. Here, $N_{\text{selected-correct}}$ represents the number of correct samples selected based on the most frequently generated patch.  
Specifically, SC identifies the patch with the highest frequency among the generated patches and verifies its correctness. By calculating $N_{\text{selected-correct}}$ (the number of correct patches) and comparing it with $N_{\text{total-correct}}$, the total number of successful repairs, SC accuracy provides an effective measure of the consistency of generated patches while also reflecting the diversity of the model's reasoning paths.
\subsection{Baselines}
The baselines used in this study include NatGen\cite{chakraborty2022natgen} and the S-C and S-C+BM25 methods from \cite{ahmed2023better}. Additionally, we compare our approach with several advanced techniques, including CoT\cite{wei2022chain}, Tree-of-Thought prompt\cite{tree-of-thought-prompting}, Plan-and-Solve\cite{wang-etal-2023-plan}, and Analogical Reasoning\cite{yasunaga2024large}, to provide a comprehensive evaluation.
\section{EXPERIMENTAL RESULTS AND ANALYSIS}
\begin{table}[htbp]
\caption{Repair results of different methods on B2Fs and B2Fm datasets}
\centering
\setlength{\tabcolsep}{0.37cm}{
\begin{tabular}{lcccc}
\toprule
\textbf{Models}         & \textbf{Methods} & \textbf{B2Fs(\%)} & \textbf{B2Fm(\%)} \\
\midrule
NatGen                 & ——                 & 23.43       & 14.93       \\
—                      & S-C                & 13.50       & 15.50       \\
—                      & S-C+BM25           & 31.80       & 21.60       \\
\midrule
\multirow{5}{*}{gpt-3.5 turbo} & CoT         & 35.50       & 18.30       \\
                               & Tree-of-Thought         & 40.00       & 22.30      \\
                               & Plan-and-Solve          & 45.40       & 26.80       \\
                               & Analogical-Reasoning    & 31.40      & 15.70      \\
                               & $\bm{T}^3$                 & \textbf{46.70}       & \textbf{28.20}       \\
\midrule
\multirow{5}{*}{gpt-4o-mini}   & CoT         & 30.10       & 22.70       \\
                               & Tree-of-Thought         & 36.20       & 23.10       \\
                               & Plan-and-Solve          & 38.40       & 26.10       \\
                               & Analogical-Reasoning    & 26.10       & 18.10       \\
                               & $\bm{T}^3$                   & \textbf{48.20}       & \textbf{32.10}       \\
\bottomrule
\end{tabular}%
}
\label{tab:model_performance}
\end{table}
\subsection{\textbf{RQ1:}Performance of the proposed approach}
The experimental results are presented in Table \ref{tab:model_performance}. Specifically, the data for S-C and S-C+BM25 are sourced from \cite{ahmed2023better}, while the data for NatGen are obtained from \cite{chakraborty2022natgen}. A detailed analysis of these results is organized as follows:  

(1) Our proposed $T^3$ method demonstrates significant improvements in repair success rates across all experimental settings. On the B2Fs dataset, $T^3$ achieves a repair rate of 46.70\%, outperforming the CoT baseline by 11.20\%. This advantage extends to the more complex B2Fm dataset, where it achieves 28.20\% success (9.90\% higher than CoT). Notably, $T^3$ maintains robust performance when deployed on different model architectures, achieving 48.20\% on B2Fs and 32.10\% on B2Fm using gpt-4o-mini. These results confirm its architectural adaptability and task-general effectiveness.

(2) Among alternative approaches, the Plan-and-Solve method achieves second-best performance with 45.40\% on B2Fs and 26.80\% on B2Fm. This relative success stems from its structured problem decomposition strategy, which enables systematic error resolution. In contrast, Analogical Reasoning shows markedly lower performance (31.40\% on B2Fs, 15.70\% on B2Fm), likely due to its dependence on potentially mismatched example retrievals. This disparity highlights the importance of task-aligned reasoning frameworks, as evidenced by CoT's intermediate performance (35.50\% on B2Fs, 18.30\% on B2Fm).

(3) The performance of models varies across datasets, which reflects the impact of task complexity on different architectures. On the B2Fs dataset, GPT-3.5-turbo performs slightly better than GPT-4o-mini (46.70\% vs 48.20\%), suggesting that smaller models may sometimes make unnecessary or excessive corrections in simpler cases. However, this pattern reverses in the more complex B2Fm dataset, where GPT-4o-mini's stronger reasoning abilities lead to better performance (32.10\% vs 28.20\%). Notably, $T^3$ consistently outperforms both models, confirming that its effectiveness is due to its structured approach rather than adjustments made for a particular model.

(4) The superior performance of $T^3$ stems from its structured dual-phase approach, which first conducts a rigorous diagnostic analysis to identify root causes before executing targeted repairs. This systematic process enhances both accuracy and effectiveness in error resolution. In contrast to single-stage alternatives, $T^3$ demonstrates a substantial advantage, achieving an average improvement of 14.90\% over baseline methods. Moreover, it maintains both high success rates (46.70\% and 48.20\% on B2Fs) and operational stability across tasks of varying complexity, further underscoring its robustness and adaptability. 



\subsection{\textbf{RQ2:}Does our proposed approach improve the diversity of model inference paths?}
\begin{table}[htbp]
\caption{SC accuracy of different methods on B2Fs and B2Fm datasets}
\begin{center}
\setlength{\tabcolsep}{0.70cm}{
\begin{tabular}{lcc}
\toprule
\textbf{Method} & \textbf{B2Fs(\%)} & \textbf{B2Fm(\%)} \\ 
\midrule
CoT & 40.85 & 54.64 \\ 
Tree-of-Thought & 43.75 & \textbf{55.16} \\ 
Plan-and-Solve & 44.27 & 49.29 \\ 
Analogical-Reasoning & 48.09 & 50.32 \\ 
$\bm{T}^3$ & \textbf{48.61} & 44.68 \\ 
\bottomrule
\end{tabular}
}
\label{tab:performance_comparison}
\end{center}
\end{table}
As indicated in Table \ref{tab:performance_comparison}, $T^3$ attains a SC accuracy of 44.68\% on the B2Fm dataset, a performance that is inferior relative to other methodologies. Nonetheless, as illustrated in Table \ref{tab:model_performance}, its repair accuracy is 28.2\%, which is notably superior to alternative approaches. This implies that, for complex tasks, the $T^3$ model employs a variety of reasoning pathways to augment overall repair efficacy, notwithstanding a compromise in consistency during patch generation.
\par
On the B2Fs dataset, the $T^3$ model achieves a repair accuracy of 46.70\% and an SC accuracy of 48.60\%. This indicates that for relatively simpler tasks, the $T^3$ model effectively balances high SC accuracy with robust repair performance.
\par
Moreover, experimental evaluations across four reasoning methods CoT, Tree-of-Thought, Plan-and-Solve and Analogical Reasoning showed that Tree-of-Thought is the only method that surpasses CoT in both SC accuracy and repair accuracy. Taking advantage of this advantage, Tree-of-Thought was adopted as the core reasoning framework for subtask inference.

\subsection{\textbf{RQ3:}Whether each component in our method contributes to improving model performance?}
\begin{table}[htbp]
\caption{Impact of $T^3$ components on repair rate}
\centering
\setlength{\tabcolsep}{0.75cm}{
\begin{tabular}{lcccc}
\toprule
\textbf{Methods} & \textbf{B2Fs(\%)} & \textbf{B2Fm(\%)} \\
\midrule
CoT & 35.50 & 18.30 \\ 
\textit{w/o} plan & 37.60 & 19.90 \\ 
\textit{w/o} cause & 37.30 & 21.00  \\ 
\bm{$T^3$} & \textbf{46.70} & \textbf{28.20} \\ 
\bottomrule
\end{tabular}
}
\label{tab1}
\end{table}
In order to investigate the impact of key components on model performance, we conduct a series of ablation experiments, with the corresponding results presented in Table \ref{tab1}. Since the retrieval module is consistently used across all control groups, we do not perform ablation on it. Additionally, because the repair rate is determined based on the final patch generated during the patch production phase, ablation was not conducted on the retrieval and patch generation stages.  

(1) In the “\textit{w/o plan}” setting, the repair rate improves compared to the baseline CoT method, with B2Fs increasing from 35.50\% to 37.60\% and B2Fm rising from 18.30\% to 19.90\%. This suggests that removing the explicit planning step does not significantly hinder the model’s ability to generate effective repairs, though the improvement remains moderate.  

(2) The results of the “\textit{w/o} cause” setting indicate that causal analysis plays an essential role in enhancing repair performance. Compared to “w/o plan,” it achieves a slightly lower B2Fs (37.30\%) but a higher B2Fm (21.00\%), demonstrating that identifying and analyzing potential causes helps the model generate more precise fixes, particularly in terms of improving correctness.  

(3) The complete $T^3$ framework outperforms all ablated variants, achieving the highest repair rates with 46.70\% in B2Fs and 28.20\% in B2Fm. These results underscore the complementary nature of planning and causal reasoning, as their integration leads to substantial gains in both repair success and correctness.  

These results demonstrate that both planning and causal analysis contribute to the model’s performance, with causal reasoning playing a particularly important role in improving repair accuracy. The superior performance of $T^3$ further validates the effectiveness of integrating these components, showcasing its advantage over traditional CoT approaches.
\subsection{\textbf{RQ4:}The impact of the number of example samples on repair accuracy}
\begin{figure}[htbp]
\centerline{\includegraphics[width=0.5\textwidth]{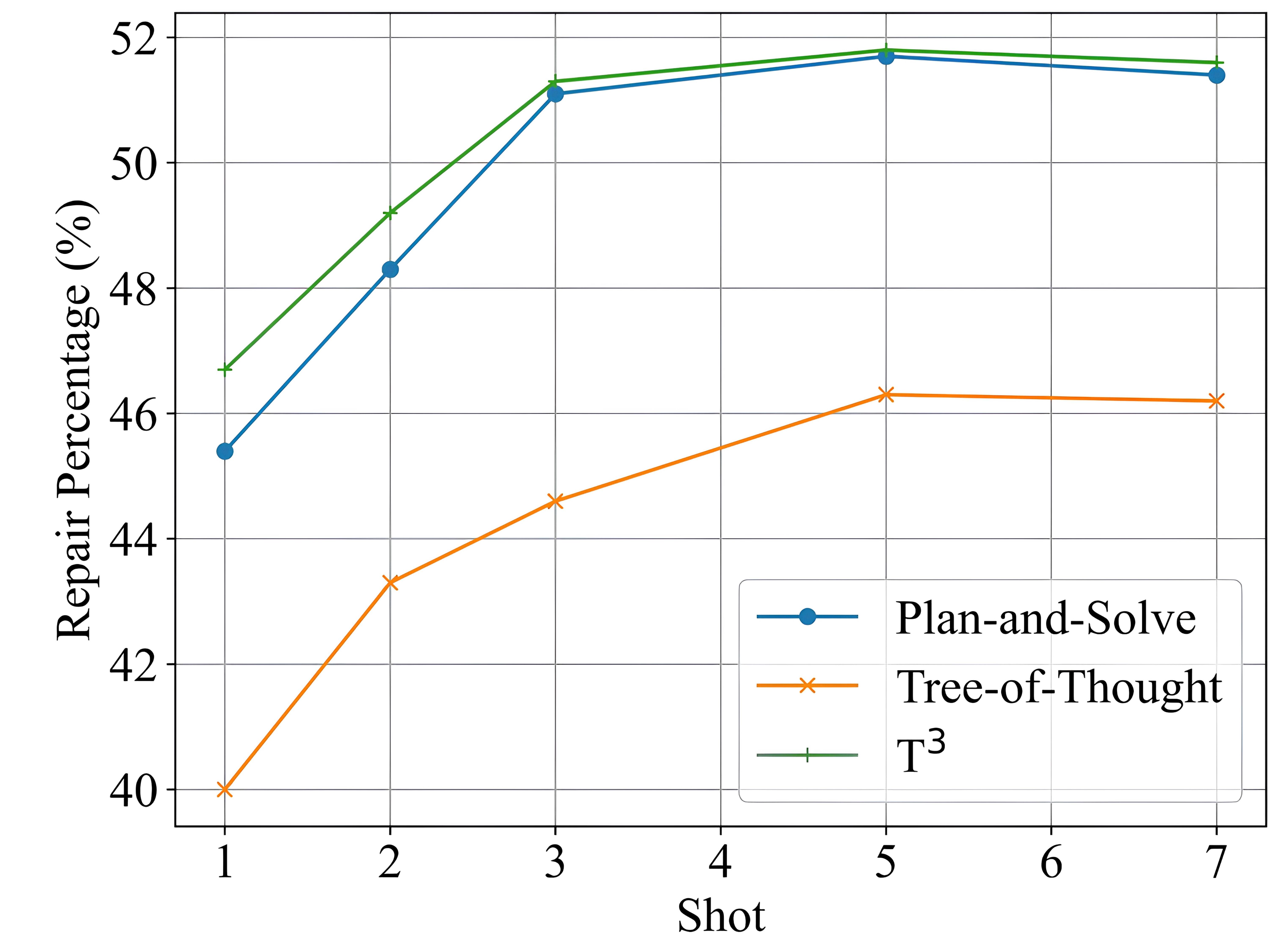}}
\caption{Repair Accuracy Performance under Different Shot Quantities}
\label{fig3}
\end{figure}
As illustrated in Fig.~\ref{fig3}, increasing the number of shots generally enhances the target repair rate. However, this improvement is not always substantial. Notably, when the number of shots reaches three, the repair rates of both the Plan-and-Solve method and the $T^3$ method stabilize, with additional shots yielding only marginal gains. In contrast, the repair rate of the Tree-of-Thought method continues to improve until reaching five shots. Nevertheless, even at five shots, its repair rate remains lower than that of the $T^3$ method with just a single shot.  
these findings provide robust empirical evidence of the $T^3$ method’s superiority in reasoning performance. Particularly in low-sample conditions, the $T^3$ method achieves a significantly higher repair rate than the Plan-and-Solve method. More importantly, as the sample size increases and both methods reach their respective stability points, the $T^3$ method continues to outperform, demonstrating not only greater effectiveness but also enhanced robustness across varying data constraints.  
Overall, the $T^3$ method not only exhibits strong generalization ability and reasoning efficiency in low-sample scenarios but also maintains consistently high performance as the sample size grows. This suggests that the $T^3$ method is highly adaptable across different sample sizes, excelling particularly in few-shot learning settings.
\subsection{Case Study}
\begin{table}[ht]
    \centering
    \caption{Comparison of repair methods for the original buggy. \textcolor{red}{Red} indicates the original error. \textcolor{blue}{Blue} denotes additional changes made during repair. \textcolor{orange}{Orange} highlights issues that remain unresolved after the revision. \textcolor{teal}{teal} marks correct repairs.}
    \label{tab:case1}
    \begin{tabularx}{\linewidth}{X}  
    \toprule
    \multicolumn{1}{c}{\textbf{Original Buggy}} \\ 
    @ java.lang.Override public void onStop () \{ super . onStop () ; if ( ( player ) != null ) \{ player . stop () ; \textcolor{red}{player . release () ; } \}\}
     \\ \midrule
    \multicolumn{1}{c}{\textbf{Repaired by CoT}} \\ 
    @java.lang.\textcolor{blue}{Overridepublic} void onStop() \{    super.onStop();    if (player != null) \{        player.stop();        \textcolor{orange}{//player . release() ; // Removed this line to fix the issue}    \}\}\\ \midrule
    \multicolumn{1}{c}{\textbf{Repaired by Tree-of-Thought}} \\
    \textcolor{blue}{private} void onStop() \{    super.onStop();    if (player != null) \{ if (player.isPlaying()) {            player.stop();} \textcolor{teal}{player.release();        player = null;}    \}\}\\ \midrule
    \multicolumn{1}{c}{\textbf{Repaired by Plan-and-Solve}} \\ 
    @ java.lang.\textcolor{blue}{Overridepublic} void onStop() \{    super.onStop();    if (player != null) \{        player.stop();   \textcolor{teal}{player.release();   player = null; // Set player to null after releasing it }   \}\}\\ \midrule
    \multicolumn{1}{c}{\textbf{Repaired by Analogical-Reasoning}} \\
    @java.lang.\textcolor{blue}{Overridepublic} void onStop() \{    super.onStop();    if ((player) != null) \{        player.stop();  \textcolor{teal}{player.release();        player = null;}    \}\}\\ \midrule
    \multicolumn{1}{c}{\textbf{Repaired by $T^3$}} \\ 
    @java.lang.Override public void onStop() \{     super.onStop();     if (player != null) \{ player.stop();     \textcolor{teal}{player . release(); player = null;}     \} \}  \\
    \bottomrule
       \end{tabularx}

\end{table}
When comparing the revised results of CoT, Tree-of-Thought, Plan-and-Solve, Analogical-Reasoning, and $T^3$ in Table \ref{tab:case1}, the advantages of $T^3$ become evident, along with the shortcomings of the other approaches. $T^3$ systematically identifies root causes and generates targeted repair plans, ensuring precise and reliable fixes while minimizing unintended modifications.  

In contrast, Plan-and-Solve follows a single reasoning path, limiting its flexibility in handling complex errors. Analogical-Reasoning relies on retrieving and adapting similar code from a knowledge base, which can be ineffective when an appropriate match is unavailable. Both methods generate repair patches in a single step, sometimes resolving the issue but often introducing unnecessary modifications to unrelated parts of the program.  

Tree-of-Thought, despite its structured approach, fails to decompose the reasoning process into clear, incremental steps. This lack of refinement reduces its effectiveness, leading to unnecessary alterations while leaving some errors unaddressed. CoT performs even less satisfactorily, as it not only fails to fix the primary error but also modifies correct parts of the target program.  

By contrast, $T^3$’s iterative approach to error diagnosis and repair ensures accuracy and efficiency, successfully addressing program errors while maintaining the program’s integrity. These distinctions underscore $T^3$’s potential as a transformative solution for program repair, overcoming challenges that existing methods fail to resolve.  

\section{Conclusion}
This paper proposes a novel APR framework, termed $T^3$, designed to leverage LLMs for handling APR tasks. The framework consists of four key stages: Sample Retrieval, Cause Analysis, Repair Plan Generation and Patch Generation.
Specifically, $T^3$ first retrieves similar code samples from a codebase based on the target program. The retrieved samples, along with the target program, are then fed into an LLM to analyze the root causes of the errors. Upon completion of the error analysis, the framework proceeds to generate a repair plan, which ultimately leads to the generation of a final patch for the target program.
Furthermore, $T^3$ investigates the correlation between the number of retrieved similar samples and the final repair accuracy. Experimental results indicate that as the number of retrieved samples increases, repair accuracy initially improves significantly before reaching a stable plateau.
The effectiveness of the proposed $T^3$ framework is demonstrated through extensive evaluations across multiple datasets and LLM configurations, including a comprehensive ablation study to validate its components and performance.

\bibliographystyle{IEEEtran}
\bibliography{APR}

\end{document}